\documentclass[11pt,british,h]{article}
\usepackage[T1]{fontenc}
\usepackage[latin1]{inputenc}
\usepackage{geometry}
\usepackage{babel}
\usepackage{graphics}
\usepackage{setspace}

\makeatletter

\providecommand{\LyX}{L\kern-.1667em\lower.25em\hbox{Y}\kern-.125emX\@}

\input epsf
\newcommand{\dir}{Figs}

\makeatother
\begin{document}
\vspace*{1cm}

{\centering \textbf{\LARGE Density functional for anisotropic fluids
}\LARGE \par}

{\centering \vspace{1cm} {\large Giorgio Cinacchi\( ^{1} \), Friederike
Schmid\( ^{2} \)}  \vspace{1cm}\par}

{\centering 1: \emph{Dipartimento di Chimica, Universita' di Pisa,
56126 Pisa, Italy} \\
 electronic mail: giorgio@giasone.dcci.unipi.it \\
 2: \emph{Fakult\"{a}t f\"{u}r Physik, Universit\"{a}t Bielefeld,
D 33501 Bielefeld, Germany}\\
 electronic mail: schmid@physik.uni-bielefeld.de\\
\par}

\vspace*{1cm}

\centerline{\textbf{\large Abstract}}

\vspace*{0.5cm}

{\small We propose a density functional for anisotropic fluids
of hard body particles. It interpolates between the well-established
geometrically based Rosenfeld functional for hard spheres and the
Onsager functional for elongated rods. We test the new approach by 
calculating the location of the the nematic-isotropic transition 
in systems of hard spherocylinders and hard ellipsoids. 
The results are compared with existing simulation data. 
Our functional predicts the location of the transition much
more accurately than the Onsager functional, and almost as good as
the theory by Parsons and Lee. We argue that it might be suited to
study inhomogeneous systems. }{\small \par}

\section{Introduction}

The density functional approach is one of the most powerful and widely
applicable approaches to nonuniform fluids~\cite{evans}. Its idea
is to express the free energy as a functional of locally varying one-particle
densities \( {\mathcal{F}}\{\rho ({\mathbf{r}})\} \), with an ideal
gas contribution and an excess free energy which accounts for the
interparticle interactions. This allows one to calculate the structure
and properties of fluids with various kinds of inhomogeneities. Density
functionals for simple fluids have reached a considerable degree of
sophistication and are able to describe such complex phenomena as
freezing, wetting, and surface melting~\cite{evans}. 

One of the most established density functionals for hard sphere fluids
is the Rosenfeld functional~\cite{rosenfeld1}. It reproduces by construction
the Percus-Yevick solution for the correlation function, which is
known to be very good~\cite{hansen}. As a consequence, it provides
highly accurate predictions for the structure of inhomogeneous hard
sphere fluids~\cite{rosenfeld2}. It has been used successfully to study
mixtures of hard spheres~\cite{rosenfeld3} and polydisperse 
fluids~\cite{pagona}. After a simple anisotropic mapping procedure 
it can be applied to fluids of fully aligned ellipsoids~\cite{rosenfeld4}. 
Exploiting the Gauss-Bonnet theorem, it has been generalized for 
isotropic molecular fluids~\cite{rosenfeld5}. This generalized functional
has been used as a starting point to calculate direct correlation
functions of isotropic multicomponent fluids~\cite{chamoux}. 
A Rosenfeld type approach has been developed for mixtures of
rods and needles, assuming that the needles are too thin to interact
with each other directly~\cite{schmidt}, and for systems containing
only one single rod immersed in a fluid of spheres~\cite{roth}.
However, to our knowledge, the Rosenfeld functional has not yet 
been extended to general anisotropic fluids. 

The simplest density functional for anisotropic particles is the Onsager
functional~\cite{onsager}, which truncates the virial expansion
after the leading coefficient. Onsager showed that this functional 
produces a nematic-isotropic phase transition~\cite{onsager} in 
fluids of sufficiently elongated particles. Its predictions are
in good agreement with experimental results on systems of tobacco
mosaic viruses~\cite{onsaesp}. One can even show that it describes 
the transition \emph{exactly} in homogeneous systems of infinitely 
elongated rods~\cite{mao}.  Thus any functional for nematic liquid 
crystals should reduce to the Onsager functional in this limit. 

In the present paper, we propose a density functional for hard anisotropic
particles which interpolates between the Rosenfeld functional and
the Onsager functional. It reduces to the Rosenfeld functional in
the case of particles with spherical symmetry, and to the Onsager
functional for homogeneous fluids of infinitely elongated particles
at density close to zero. As a first test of the functional, we have
calculated the nematic-isotropic transition for hard spherocylinders
and hard ellipsoids, and obtained reasonable results. We believe that our
functional might provide a useful new approach to the study of inhomogeneous
liquid crystals, e.g. the study of interfacial phenomena such as surface
anchoring.

\section{Background}

We consider a fluid of hard anisotropic particles with positions 
\( {\mathbf{r}} \) and orientations \( \varpi  \). 
The grand canonical free energy is
the minimum of the free energy functional 
\begin{equation}
\label{1}
\beta \Omega [\rho ] = \beta {\mathcal{F}}^{id}[\rho ] 
+ \beta {\mathcal{F}}^{ex}[\rho ] + 
\int d{\mathbf{r}}\: d\varpi \: \rho ({\mathbf{r}},\varpi )\: 
[\mu -V({\mathbf{r}},\varpi )]
\end{equation}
 with respect to the one-particle density \( \rho ({\mathbf{r}},\varpi ) \),
where \( \beta =\frac{1}{k_{B}T} \) is the Boltzmann factor with
the temperature \( T \), \( \mu  \) the chemical potential and 
\( V({\mathbf{r}},\varpi ) \) summarises all external potentials. 
The first term describes the ideal
gas contribution 
\begin{equation}
\label{p}
\beta {\mathcal{F}}^{id} = 
\int d{\mathbf{r}}\: d\varpi \: \rho ({\mathbf{r}},\varpi )\: 
\Big (\ln [\rho ({\mathbf{r}},\varpi )\lambda _{T}^{3}]-1\Big )
\end{equation}
 with the de Broglie wavelength \( \lambda _{T} \). The second term
\( {\mathcal{F}}^{ex} \) is the excess free energy, the central quantity
in density functional theories~\cite{evans}. 

In the Onsager theory, the excess free energy of two hard particles
is given by \begin{equation}
\label{onsager}
\beta {\mathcal{F}}_{O}^{ex} = - \frac{1}{2}
\int d{\mathbf{r}}\: d\varpi \: d{\mathbf{r}}'\: d\varpi' \: 
  \rho ({\mathbf{r}},\varpi )\: \rho ({\mathbf{r}}',\varpi' )\: 
   \mbox {M}({\mathbf{r}}-{\mathbf{r}}',\varpi ,\varpi' ).
\end{equation}
 The Mayer-function {\mbox {M}} takes the value \( (-1) \) if two particles
at \( ({\mathbf{r}},\varpi ) \) and \( ({\mathbf{r}}',\varpi' ) \)
overlap and vanishes otherwise. In the homogeneous case, the one-particle
density \( \rho  \) does not depend on \( {\mathbf{r}} \) and can
be written as \( \rho ({\mathbf{r}},\varpi )=\rho \: f(\varpi ) \).
The expression \ref{onsager} then reduces to \begin{equation}
\label{6}
\frac{\beta {\mathcal{F}}_{O}^{ex}}{N} = \frac{\rho }{2}
\int d\varpi \: d\varpi' \: f(\varpi )\: f(\varpi' )\: \nu (\varpi ,\varpi' ),
\end{equation}
 where \( \nu (\varpi ,\varpi ') \) is the covolume of two particles
with orientations \( \varpi  \) and \( \varpi'  \), and \( N \)
the total number of particles. Eqn.\ (\ref{6}) corresponds to a virial
expansion up to second order. As mentioned in the introduction, the
Onsager theory is exact in the limit of infinitely elongated particles
~\cite{mao}. Compared to real thermotropic liquid crystals, it tends
to overestimate the shape anisotropy required to observe a nematic
phase at a given finite density. 

A natural extension of the Onsager model would be to include higher
order terms in the virial expansion. On principle, this is feasible,
and an extension of the Onsager functional up to third order was actually
carried out for ellipsoids ~\cite{tiptomargo}. However , the calculations
are very cumbersome due to the increasing complexity of the integrals. 
As long as one is interested in homogeneous systems, one way to overcome 
the problem is the \emph{decoupling approximation.}
It consists of a resummation of the virial expansion, where the
first virial coefficients are calculated exactly, and the remaining
ones are approximated by a mapping onto a reference system. Given
a virial expansion of the form:
\begin{equation}
\label{virialex}
\frac{\beta \mathcal{F}^{ex}}{N} =
\sum ^{\infty }_{l=1}
\frac{\rho ^{l}}{l}B_{l+1}\left[ f\left( \varpi \right) \right] ,
\end{equation}
the decoupling approximation reads~\cite{parsons}-\cite{padilla}:
\begin{equation}
\label{DA}
\frac{\beta \mathcal{F}^{ex}}{N} =
\sum ^{r}_{l=1}\frac{\rho ^{l}}{l}B_{l+1}\left[ f\left( \varpi \right) \right] 
+ \left[ \frac{\beta F^{ex,ref}}{N}
    - \sum ^{r}_{l=1}\frac{\rho ^{l}}{l}B^{ref}_{l+1} \right] 
\frac{B_{r+1}\left[ f\left( \varpi \right) \right] }{B^{ref}_{r+1}},
\end{equation}
where \( B_{l+1}\left[ f\left( \varpi \right) \right]  \) are the
virial coefficients of the anisotropic fluids, calculated exactly
up to the \( r+1 \) order and {\it ref } refers to the reference
system. Two possible reference systems have been proposed: the hard
sphere fluid and the isotropic fluid of the particles under investigation.
If \( r=1 \) and the reference system is the hard sphere fluid with
the Carnahan-Starling equation of state~\cite{carnahan}, 
we recover the Parsons-Lee functional~\cite{parsons,lee} 
\begin{equation}
\label{parsons}
\beta {\mathcal{F}}_{PL}^{ex} = -
  \frac{1}{8} \frac{(4-3\eta )}{(1-\eta )^{2}}
  \int d{\mathbf{r}} \: d\varpi \: d{\mathbf{r'}} \: d\varpi' 
    \: \rho({\mathbf{r}},\varpi ) \: \rho({\mathbf{r'}},\varpi') \: 
  \mbox M({\mathbf{r}}-{\mathbf{r'}},\varpi ,\varpi' ),
\end{equation}
where \( \eta  \) is the packing fraction. The Parsons-Lee functional
predicts accurately the location of the nematic-isotropic phase transition
for a wide range of shape anisotropies~\cite{mcgrother,camp}. 
It can also be applied to inhomogeneous systems. However,
it does not describe the microscopic structure very well: it yields
very crude correlation functions~\cite{allen,note} and produces
rather unrealistic density profiles in inhomogeneous systems as a
consequence~\cite{phuong1}. 

The success of the Parsons-Lee theory in predicting the isotropic-nematic 
coexistence densities motivated Poniewierski and Holyst~\cite{Holyst}
and Somoza and Tarazona~\cite{ST1} to combine the decoupling approximation
with a ``weighted density approximation'' (WDA) 
scheme~\cite{WDATarazona,Tarazona,WDACurtin}. 
Both are  constructed such that one automatically recovers the Onsager 
functional in the low density limit. In the hard sphere limit,
Somoza's and Tarazona's~\cite{ST1} and related~\cite{Velasco} versions
reduce to the well-known Tarazona functional~\cite{Tarazona}. The
latter incorporates the Carnahan-Starling equation of state and the 
Percus-Yevick direct correlation function for homogeneous hard sphere 
fluids, and has been used with great success to study inhomogeneous 
hard sphere fluids and solids. Applied to systems of spherocylinders,
the extensions~\cite{ST1,Velasco} of the Somoza functional generate a 
nematic phase and several smectic phases. Graf and L\"owen have put
forward a simplified ``modified weighted density approximation'' (MWDA),
and used it to reproduce complex phase diagrams of spherocylinder fluids~\cite{Lowen}.

Besides the Tarazona functional, another equally successful hard sphere 
functional has established itself in recent years: the fundamental measure 
theory or Rosenfeld functional~\cite{rosenfeld1,rosenfeld2,rosenfeld3}. 
It has the advantage of being based on somewhat more fundamental
considerations: it does not require the explicit input of the 
equation of state and the direct correlation functions, instead
they pop out automatically. One obtains by construction the
Percus-Yevick solution. Moreover, the functional is formulated
a priori for general convex particles of arbitrary geometry.
It thus seems to call for a generalization to anisotropic fluids. 
Unfortunately, there is one problem: the original Rosenfeld 
functional does not reproduce the Onsager functional in the 
low density limit. 

Here, we propose a simple modification of the Rosenfeld functional
which solves that problem for anisotropic particles, and reduces 
to the original Rosenfeld functional for isotropic particles. 
Like Somoza's and Tarazona's functional, our functional 
interpolates between an established hard sphere functional
and the Onsager functional. Thus we hope that it will be
equally useful.

Before introducing our approach, we sketch briefly the basic
equations of the original Rosenfeld functional~\cite{rosenfeld1}.
It can be formulated for general multicomponent fluids of convex hard
particles. In our case, different components \( i \) may be identified
with different particle orientations. The functional then reads:
\begin{equation}
\label{8bis}
\beta {\mathcal{F}}_{R}^{ex} =
\int d{\mathbf{r}}\: \Big (\: \Phi _{1}({\mathbf{r}})
+ \Phi _{2}({\mathbf{r}})+\Phi _{3}({\mathbf{r}})\: \Big ),
\end{equation}
\begin{equation}
\label{8.1bis}
\Phi _{1}({\mathbf{r}})=-n_{0}({\mathbf{r}})\ln (1-n_{3}({\mathbf{r}}))
\end{equation}
\begin{equation}
\label{8.2bis}
\Phi _{2}({\mathbf{r}}) =
\frac{n_{1}({\mathbf{r}})n_{2}({\mathbf{r}})
- {\mathbf{n}}_{1}({\mathbf{r}})\cdot 
{\mathbf{n}}_{2}({\mathbf{r}})}{1-n_{3}({\mathbf{r}})}
\end{equation}
\begin{equation}
\label{8.3bis}
\Phi _{3}({\mathbf{r}}) =
\frac{\frac{1}{3}n_{2}^{3} ({\mathbf{r}})
     - n_{2}({\mathbf{r}})({\mathbf{n}}_{2}({\mathbf{r}})\cdot 
          {\mathbf{n}}_{2}({\mathbf{r}}))}
     {8\pi (1-n_{3}({\mathbf{r}}))^{2}} \cdot \lambda
\end{equation}
 The \( \Phi  \)s depend on the weighted densities \begin{equation}
\label{12}
n_{\alpha }({\mathbf{r}}) =
\sum _{i}\int d{\mathbf{r}}^{'}\rho _{i}({\mathbf{r}}^{'})\: 
     w^{(\alpha )}_{i}({\mathbf{r}}-{\mathbf{r}}^{'}),
\end{equation}
with the number density of the \( i \)th component 
\( \rho _{i}({\mathbf{r}}) \) and the weight functions 
\begin{equation}
\begin{array}{rcl  rcl}
w^{(0)}_{i}({\mathbf{r}}) 
& = &
C_i({\hat{\mathbf{r}}}) \:
\delta (R_{i}({\hat{\mathbf{r}}})-r)\: K_{i}({\hat{\mathbf{r}}})/4\pi 
&
w^{(2)}_{i}({\mathbf{r}}) 
& = &
C_i({\hat{\mathbf{r}}}) \:
\delta (R_{i}({\hat{\mathbf{r}}})-r) 
\\
w^{(1)}_{i}({\mathbf{r}}) 
& = &
C_i({\hat{\mathbf{r}}}) \:
\delta (R_{i}({\hat{\mathbf{r}}})-r)\: H_{i}({\hat{\mathbf{r}}})/4\pi 
&
w^{(3)}_{i}({\mathbf{r}}) 
& =  &
\Theta (R_{i}({\hat{\mathbf{r}}})-r)\label{weight} 
\\
{\mathbf{w}}^{(1)}_{i}({\mathbf{r}}) 
& = &
C_i({\hat{\mathbf{r}}}) \:
 \delta (R_{i}({\hat{\mathbf{r}}})-r)\: 
    {\hat{\mathbf{n}}}_{i}({\hat{\mathbf{r}}})\: H_{i}({\hat{\mathbf{r}}})/4\pi 
\qquad 
&
{\mathbf{w}}^{(2)}_{i}({\mathbf{r}}) 
& = & 
C_i({\hat{\mathbf{r}}}) \:
\delta (R_{i}({\hat{\mathbf{r}}})-r)\:
    {\hat{\mathbf{n}}}_{i}({\hat{\mathbf{r}}}).
\end{array}
\end{equation}
Here \( {\hat{\mathbf{r}}} \) denotes the unit vector in the direction
of \( {\mathbf{r}} \), \( R_{i}({\hat{\mathbf{r}}}) \) the radius
from the centre of a particle of type \( i \) to the surface in the
direction \( {\hat{\mathbf{r}}} \), \( H_{i}({\hat{\mathbf{r}}}) \)
and \( K_{i}({\hat{\mathbf{r}}}) \) the local mean and Gaussian curvature
on that specific surface point~\cite{note2,weatherburn}, 
\( {\hat{\mathbf{n}}}_{i}({\hat{\mathbf{r}}}) \)
the outward unit normal (see Figure 1), and \( \delta  \) and \( \theta  \)
are the usual delta and step function. The factor $C_i({\hat{\mathbf{r}}})$
ensures that integrals over the weight functions 
$w^{(\alpha)}_{i}({\mathbf{r}})$ and 
${\mathbf{w}}^{(\alpha)}_{i}({\mathbf{r}})$
$(\alpha \ne 3)$ are surface integrals over the surface of the particle $i$.
For spherical particles, it can be omitted. In general, it is given by
\begin{equation}
C_i({\hat{\mathbf{r}}}) = \sqrt{\det \mathbf{g}_i({\hat{\mathbf{r}}})/
                              \det \mathbf{g}_{i,\mbox{sph}}({\hat{\mathbf{r}}}) },
\end{equation}
where $\mathbf{g}_i({\hat{\mathbf{r}}})$ is the local metric tensor of
the particle $i$, and $\mathbf{g}_{i,\mbox{sph}}({\hat{\mathbf{r}}})$ that of a reference
sphere of radius $R_{i}({\hat{\mathbf{r}}})$. (In polar coordinates,
$\sqrt{\det \mathbf{g}_{i,\mbox{sph}}({\hat{\mathbf{r}}})} = R_i^2({\hat{\mathbf{r}}})
\sin(\theta)$). 
Finally, the factor $\lambda$ in Eqn. (\ref{8.3bis}) ensures that the dimensional
crossover for hard spheres is reproduced correctly and that the hard sphere
system exhibits a solid-fluid transition in three dimensions~\cite{rosenfeld2}.
\begin{equation}
\label{csi}
\lambda =
 \frac{\left( 1-\xi ({\mathbf{r}})^{2}\right) ^{3}}
           {1-3\xi ({\mathbf{r}})^{2}}
\quad \mbox{with} \quad
\xi =\left| \frac{{\mathbf{n}}_{2}({\mathbf{r}})}{n_{2}({\mathbf{r}})}\right|.
\end{equation}
Alternatively, Eqn. (\ref{8.3bis}) can be replaced by a more modern version 
due to Tarazona~\cite{tensore}, which depends on an additional tensorial
weighted density and does not contain the {\em ad hoc} factor $\lambda$.

\begin{figure}
{\centering \resizebox*{0.3\textwidth}{!}{\includegraphics{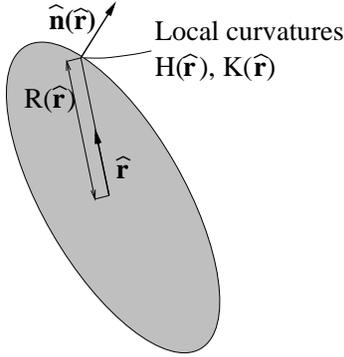}} \par}

\caption{Illustration of \protect\( R(\widehat{\mathbf{r}})\protect \) and
\protect\( \widehat{\mathbf{n}}(\widehat{\mathbf{r}})\protect \).}
\end{figure}

\vspace{1cm}

\section{Construction of the functional}

As mentioned before, the Rosenfeld functional describes hard sphere fluids 
very successfully. On principle, it could also be 
applied to molecular fluids. However, it has the serious drawback
that it does not reduce to the Onsager functional in the low
density limit. Even worse, a closer inspection of 
Eqns. (\ref{8bis})--(\ref{csi}) reveals that it has no contribution
{\em at all} which would favor parallel alignment of particles in 
a homogeneous fluid. Hence it cannot produce stable homogeneous 
anisotropic fluids. 

The reason for this failure can be understood by looking at the relation
between the Mayer-function and the weight functions in more detail.
For hard spheres, the Mayer-function for a pair of particles at position
\( {\mathbf{r}}_{i} \) and \( {\mathbf{r}}_{j} \) can be decomposed
exactly as~\cite{rosenfeld1}
\begin{equation}
\label{deconvolution}
-\mbox {M}_{ij} =
w_{i}^{(0)}\otimes w_{j}^{(3)} + 
w_{i}^{(3)}\otimes w_{j}^{(0)} +
w_{i}^{(1)}\otimes w_{j}^{(2)} +
w_{i}^{(2)}\otimes w_{j}^{(1)} -
{\mathbf{w}}_{i}^{(1)}\otimes {\mathbf{w}}_{j}^{(2)} -
{\mathbf{w}}_{i}^{(2)}\otimes {\mathbf{w}}_{j}^{(1)},
\end{equation}
 where \( \otimes  \) denotes the convolution product: 
\begin{equation}
w_{i}^{(\alpha )}\otimes w_{j}^{(\beta )} =
\int d{\mathbf{r}}\: w^{(\alpha )}({\mathbf{r}}-{\mathbf{r}}_{i})\: 
w^{(\beta )}({\mathbf{r}}-{\mathbf{r}}_{j}).
\end{equation}
 This decomposition together with Eqns.\ (\ref{8bis})--(\ref{12}) ensures
that the functional reproduces the correct virial expansion at least
up to second order. Unfortunately, a decomposition of the type above
is no longer valid for anisotropic particles. Therefore, the Onsager
limit is not recovered. Chamoux and Perera~\cite{chamoux} have proposed
ways to cure the virial expansion on the level of the direct correlation 
function, for the case of isotropic molecular fluids~\cite{chamoux}. 
A systematic way of dealing with the problem on the level of the functional 
itself would be to add a correction term on the right hand side of 
Eqn.~(\ref{deconvolution}), to deconvolute it (if possible), and then 
rederive a Rosenfeld type functional on the basis of this new decomposition. 
This can be done for rod-sphere interactions~\cite{roth}. 
An extension to the general case is currently under way~\cite{mecke}. 
Unfortunately, it turns out that infinitely many additional terms are 
required in the decomposition in order to make Eqn.~(\ref{deconvolution}) 
exact, and the numerical treatment becomes difficult unless one resorts 
to approximations.

Here we propose a simpler, numerically perhaps more tractable Ansatz --
a straightforward modification of Eqns. (\ref{8.1bis}), (\ref{8.2bis}) and 
(\ref{8.3bis}).  Instead of keeping \( n_{1} \) and \( {\mathbf{n}}_{1} \) 
as separate quantities, we suggest to replace the products 
\( n_{1}({\mathbf{r}})n_{2}({\mathbf{r}}) \) and 
\( {\mathbf{n}}_{1}({\mathbf{r}})\cdot {\mathbf{n}}_{2}({\mathbf{r}}) \)
by new joint quantities \(n_{12}({\mathbf{r}}) \) and 
\(\tilde{n}_{12}({\mathbf{r}}) \) such that the functional reproduces
correctly the Onsager limit. This is achieved as follows:
we define the functions
\begin{eqnarray}
g(\varpi_i,\varpi_j,{\mathbf{r}}_i - {\mathbf{r}}_j)
\equiv g_{ij} & := &
-M_{ij}
-w_{i}^{(0)}\otimes w_{j}^{(3)} 
-w_{i}^{(3)}\otimes w_{j}^{(0)}  \\
h(\varpi_i,\varpi_j,{\mathbf{r}}_i - {\mathbf{r}}_j)
\equiv h_{ij} & := &
w_{i}^{(1)}\otimes w_{j}^{(2)} +
w_{i}^{(2)}\otimes w_{j}^{(1)} -
{\mathbf{w}}_{i}^{(1)}\otimes {\mathbf{w}}_{j}^{(2)} -
{\mathbf{w}}_{i}^{(2)}\otimes {\mathbf{w}}_{j}^{(1)},
\end{eqnarray}
and
\begin{equation}
\label{heff}
H^{\mbox {\small eff}}
 (\varpi_{i},\varpi_{j},{\mathbf{r}}_{i}-{\mathbf{r}}_{j})
= g_{ij}/h_{ij}.
\end{equation}
Then we replace \( n_{1}({\mathbf{r}})n_{2}({\mathbf{r}}) \) and 
\( {\mathbf{n}}_{1}({\mathbf{r}})\cdot {\mathbf{n}}_{2}({\mathbf{r}}) \)
by

\parbox{\textwidth}{
\begin{eqnarray}
\label{13}
n_{12}({\mathbf{r}}) &:=&
\frac{1}{2} \int \! \! d\varpi_{i}\: d\varpi_{j}\: 
  d{\mathbf{r}}_{i}\: d{\mathbf{r}}_{j}\: 
  \rho ({\mathbf{r}}_{i},\varpi _{i})\: 
  \rho ({\mathbf{r}}_{j},\varpi _{j})\: 
  H^{\mbox {\small eff}}
    (\varpi _{i},\varpi _{j},{\mathbf{r}}_{i} - {\mathbf{r}}_{j}) \\
&& \times  
  \Big( w^{(1)}_{i}({\mathbf{r}}_{i}-{\mathbf{r}})\: 
        w^{(2)}_{j}({\mathbf{r}}_{j}-{\mathbf{r}})
 + 
        w^{(2)}_{i}({\mathbf{r}}_{i}-{\mathbf{r}})\: 
        w^{(1)}_{j}({\mathbf{r}}_{j}-{\mathbf{r}}) \Big)
  \nonumber
\\
\label{14}
\tilde{n}_{12}({\mathbf{r}}) &:= & 
\frac{1}{2} \int \! \! d\varpi _{i}\: d\varpi _{j}\: 
  d{\mathbf{r}}_{i}\: d{\mathbf{r}}_{j}\: 
  \rho ({\mathbf{r}}_{i},\varpi_{i})\: 
  \rho ({\mathbf{r}}_{j},\varpi _{j})\: 
  H^{\mbox {\small eff}}
    (\varpi _{i},\varpi _{j},{\mathbf{r}}_{i} - {\mathbf{r}}_{j}) \\
&& \times
 \Big( {\mathbf{w}}^{(1)}_{i}({\mathbf{r}}_{i} - {\mathbf{r}})
 \cdot {\mathbf{w}}^{(2)}_{j}({\mathbf{r}}_{j} - {\mathbf{r}})\: 
+
       {\mathbf{w}}^{(2)}_{i}({\mathbf{r}}_{i} - {\mathbf{r}})
 \cdot {\mathbf{w}}^{(1)}_{j}({\mathbf{r}}_{j} - {\mathbf{r}}) \Big).
\nonumber
\end{eqnarray}
}
 The effective function \( H^{\mbox {\small eff}} \) is constructed
 such that the functional reduces to the Onsager limit
(\ref{onsager}) in the low density limit. Thus the mean curvature
\( H({\hat{\mathbf{r}}}) \) in the weight functions \( w^{(1)} \)
and \( {\textbf {w}}^{(1)} \) (Eqn.\ (\ref{weight})) is effectively
replaced by a function \( H\cdot H^{\mbox {\small eff}} \) which
depends on the orientations of a pair of particles, and their distance
vector. In the hard sphere limit, \( H^{\mbox {\small eff}} \) is
constant, \( H^{\mbox {\small eff}}\equiv 1 \), and we recover the
original Rosenfeld functional. Note that $n_{12}$ and 
$\tilde{n}_{12}$ contain information on {\em pairs} of interacting
particles. We thus give up the idea of formulating a functional
which depends only on single, orientation independent weighted 
densities~\cite{footnote}. An approach in the same spirit has been 
introduced by Schmidt for mixtures of spheres and needles~\cite{schmidt}.

The functional can be simplified by making the approximation that
\( H^{\mbox {\small eff}} \) only depends on the orientations 
\( \varpi _{i} \) and \( \varpi _{j} \). In that case we solely require 
that the second virial coefficient is exact in the \emph{homogeneous} 
fluid for every distribution of orientations. In the homogeneous fluid 
the contributions to the free energy density having vectorial character 
vanish. Carrying out the integrals on the spatial variables, we obtain 
\begin{equation}
\label{15}
\frac{\beta {\mathcal{F}}_{R}^{ex}}{N} = 
-\ln \left( 1-\eta \right)  +
\frac{\rho \: \overline{R}\: S
  \int \! \! d\varpi \: d\varpi' \: f(\varpi )\: f(\varpi ')\: 
  H^{\mbox {\small eff}}(\varpi ,\varpi' )}{1-\eta } +
  \frac{1}{24\pi }\frac{\rho ^{2}S^{3}}{(1-\eta )^{2}},
\end{equation}
 where \( \overline{R} \), \( S \) are the mean radius and the surface
of the body, respectively, and \( \eta  \) is the packing fraction.
The mean radius is defined as the integral of the mean curvature over
the surface of the particle, \( \overline{R}=\int _{S}dA\: H/4\pi  \).
Eqn.\ (\ref{15}) must reduce to the Onsager functional (\ref{6})
in the low density limit \( \rho \rightarrow 0 \). With \( \eta =v_{0}\rho  \)
this leads to the equation 
\begin{equation}
\label{heff2}
H^{\mbox {\small eff}}(\varpi ,\varpi ')=\frac{\frac{1}{2}
\nu (\varpi,\varpi' )-v_{0}}{\overline{R}S},
\end{equation}
which replaces Eqn.~(\ref{heff}) in this approximation.

In the case of homogeneous fluids, both (\ref{heff}) or (\ref{heff2}) 
give the same bulk free energy as a function the orientation distribution
$f(\varpi)$: 
\begin{equation}
\label{final}
\frac{\beta {\mathcal{F}}_{R}^{ex}}{N} = 
  -\ln \left( 1-\eta \right) 
  -\frac{\eta }{1-\eta }
  +\frac{\rho }{2}\frac{
    \int \! \! \varpi \: d\varpi' \: f(\varpi )\: f(\varpi' )\: 
     \nu (\varpi ,\varpi' )}{1-\eta }
  +\frac{1}{24\pi }\frac{\rho ^{2}S^{3}}{(1-\eta )^{2}}
\end{equation}
Note that for isotropic fluids, $f(\varpi)$ is constant, 
and the free energy (\ref{final}) is identical to that obtained with
the original Rosenfeld functional.

\section{Application to spherocylinders and ellipsoids}

We have tested our approach by calculating the location of the 
nematic-isotropic phase transition for hard spherocylinders and
uniaxial prolate ellipsoids. The results were compared with the 
corresponding phase diagrams obtained from the Onsager theory,
the Parsons-Lee theory, and simulation data~\cite{camp,bolhuis,frenkell}.

A spherocylinder consists of a cylinder of length \( L \) and diameter
\( D \) capped by a hemisphere of diameter \( D \) at both ends
(see Figure 2).

The excluded volume of two spherocylinders which have the angle \( \theta  \)
with respect to each other is~\cite{onsager} \begin{equation}
\label{covosfe}
\nu (\theta )=\: 2DL^{2}\sin (\theta )+2\pi D^{2}L+4\pi D^{3}/3.
\end{equation}

\begin{figure}
{\centering \resizebox*{0.5\textwidth}{!}{\includegraphics{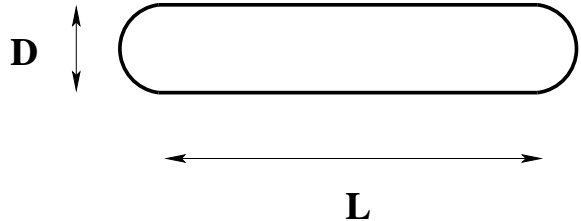}} \par}

\caption{{\small Illustration of a spherocylinder}}
\end{figure}

In the case of ellipsoids, the exact calculation of the excluded
volume is quite involved. One possibility is to follow Ref.~\cite{isihara}, 
another to exploit the Perram and Wertheim routine for the ellipsoids 
contact function~\cite{perram}. Here, we have adopted a scheme outlined 
in Ref. ~\cite{camp}, because it is sensibly faster. 
We consider a pair of equal uniaxial prolate ellipsoids
with semiaxes \( a \), \( a \) and \( b \); one ellipsoid is mapped
onto a sphere of radius \( a \), and the other one is mapped onto
a particular biaxial ellipsoid which has always one semiaxis equal
to \( a \). The remaining two semiaxes depend on the relative orientation
of the two ellipsoids and can easily be evaluated. The excluded volume
between the sphere and the biaxial ellipsoids can be calculated using
an expression by Kihara for the orientationally averaged excluded volume 
between two convex hard particles ~\cite{kihara}. The covolume of the
original pair of uniaxial prolate ellipsoids is then recovered by
inverting the mapping. In the past, many theoretical 
studies~\cite{parsons,lee,Singh,Colot,Marko} have considered systems 
of ``hard Gaussian overlap'' particles, i. e., ellipsoids where the 
contact distance is approximated by an expression due to 
Berne and Pechukas ~\cite{pechukas}.
However, it has been noted some years ago~\cite{tiptomargo}
and shown recently~\cite{miguel} that the comparison of
these theoretical results with true ellipsoids behaviour is not appropriate. 

Both in the cases of hard spherocylinders and hard ellipsoids, the
properties of the fluids are fully determined by the density and the
shape anisotropy parameters, \( x=L/D \) and \( k=b/a \), respectively.

In the homogeneous fluid, the orientation dependent part of the excess
free energy has the same form in the Onsager model, the Parsons-Lee
model, and our model: 
\begin{equation}
\label{prima}
\frac{\beta {\mathcal{F}}^{ex}[f(\varpi )]}{N} = \frac{\lambda }{2}
 \int d\varpi \: d\varpi' \: f(\varpi )\: f(\varpi' )\: \nu (\varpi ,\varpi '),
\end{equation}
where $\lambda =\rho$ in the Onsager model, 
$\lambda = \frac{\rho}{8}\frac{ (4-3\eta )}{(1-\eta )^{2}}$ in the
Parsons-Lee model, and $\lambda =\frac{\rho}{2(1-\eta)}$ in our model.
 Once a model free energy functional is chosen, the next step is to
determine the thermodynamically stable phase for a range of densities.
We thus need to calculate the functions \( f(\varpi ) \) that extremize
the free energy under the constraint \( \int \!\! d\varpi \: f(\varpi )=1 \).
For all three models under consideration, this amounts to finding
the solution of the Onsager integral equation \begin{equation}
\label{24}
\ln (\xi \: f(\varpi )) =
-\lambda \int d\varpi' \: f(\varpi ')\: \nu (\varpi ,\varpi' ),
\end{equation}
 where the constant \( \xi  \) is determined from the normalisation
condition. The equation always has an isotropic solution  
\( f(\varpi )= constant\), and may have an anisotropic solution 
in addition at certain values of \( \lambda  \). 

\begin{figure}
{\centering \resizebox*{0.6\textwidth}{!}{\includegraphics{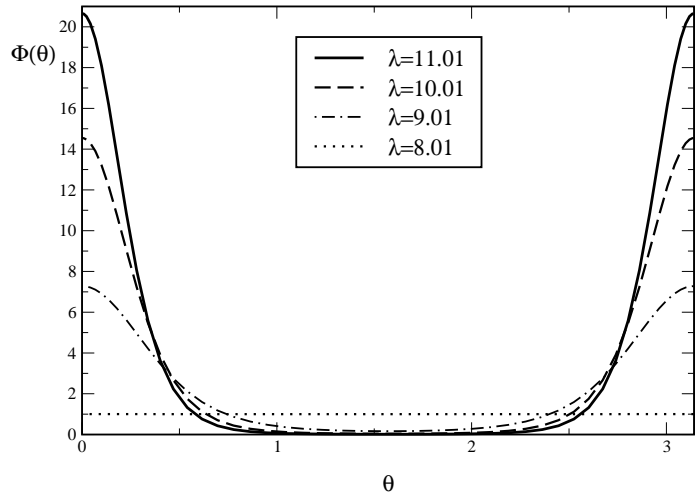}} \par}

\caption{{\small Solutions \protect\( \Phi (\theta )=4\pi f(\theta )\protect \)
of the Onsager integral equation for spherocylinders for 
\protect\( \lambda ^{*}=11.01\protect \),
10.01, 9.01, 8.01. }}
\end{figure}

We have solved the Onsager integral equation for values of 
\( \lambda  \) in the range of 
\( \lambda ^{*}=2DL^{2}\lambda\in [8:12] \) for spherocylinders, and 
\( \lambda ^{*}=2ab^{2}\lambda  \in [8:25] \)
for ellipsoids, in steps of 0.01. To this end, we have applied an iterative 
numerical method, which is simple and reliably convergent~\cite{herzfeld}.
The integrals were calculated by the Gauss-Legendre quadrature, using 
50 points per integral and checking that with 100 points we get the same 
results.  The initial guess for the highest value of \(\lambda^*\),  
\( \lambda ^{*}=12 \) or \( \lambda ^{*}=25 \), was 
\[
f(\theta ,\phi ) = \frac{1}{2\pi }\frac{\exp 
\left( P_{2}(\cos \theta )\right) }{\int du\: \exp \left( P_{2}(u)\right) },
\]
 where \( P_{2}(u) \) is the second Legendre polynomial. At lower
\( \lambda ^{*} \), the iteration was started with the anisotropic 
solution  for the previous, next higher, value 
of \( \lambda ^{*} \). The convergence criterion was
\begin {equation}
\label {converge}
\left [ \sum^{n_{p}}_{i=1}\left 
[f_{m+1}(\theta_{i})-f_{m}(\theta_{i})\right ]^{2} \right]^{\frac{1}{2}} < 10^{-6}  
\end{equation}
where \emph {$n_{p}$} refers to the number of points used to perform integrals 
and \emph {m} to the \emph {m}th iteration. 
The dependence on  \( \phi  \)  has been omitted because 
the distribution function does not actually depend on it. 
The resulting distribution functions for \( \lambda ^{*}=11.01 \),
10.01, 9.01, 8.01 for spherocylinders are shown in Figure 3. 
Our orientational distribution functions compare well with those 
in reference ~\cite{roijmulder}. Due to the particular shape of 
the covolume between a pair of spherocylinders, the solutions 
depend on \( \lambda ^{*} \), but not on \( x \).
We found anisotropic solutions for values \( \lambda ^{*}\geq 8.88 \),
in agreement with Lasher~\cite{lasher}.   

\begin{figure}
{\centering \resizebox*{0.6\textwidth}{!}{\includegraphics{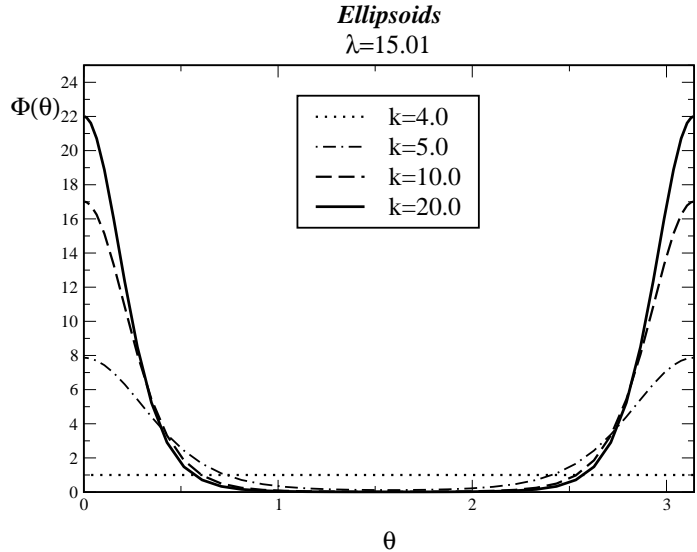}} \par}

\caption{{\small Solutions 
\protect\( \Phi (\theta )=4\pi f(\theta )\protect \)
of the Onsager integral equation for ellipsoids for 
\protect\( k=4.0\protect \),
5.0, 10.0, 20.0 at fixed value of \protect\( \lambda ^{*}=15.01\protect \).}}
\end{figure}
 
In the case of ellipsoids, the solution for fixed \( \lambda ^{*} \) depend
on the shape anisotropy parameter \( k \). In Figure 4, solutions
at different value of \( k \) are shown for \( \lambda ^{*}=15.01 \).
The lowest value of \( \lambda ^{*} \) which still yields an anisotropic 
solution is a decreasing function of \( k \).
 
Next we must identify the stable phases and the coexistence line.
At coexistence, both the pressure and the chemical potential are equal:
\begin{equation}
P_{N}=P_{i};\qquad \qquad \mu _{N}=\mu _{i}.
\end{equation}
 We have solved these two equations with Newton's method, i.e., we
found the density \( \rho ^{P}_{I} \) at which an isotropic phase
has the pressure \( P_{N} \) and the density \( \rho ^{\mu }_{I} \)
at which an isotropic phase has the chemical potential equal to \( \mu _{N} \).
At the coexistence line, both densities must be equal. Hence we calculated
the nematic \( \rho _{N} \) at which \( |\rho ^{P}_{I}-\rho ^{\mu }_{I}| \)
was minimal. Usually the value of the minimum was in the 
range \( |\rho ^{P}_{I}-\rho ^{\mu }_{I}|\in [10^{-9};10^{-5}] \). 

The results are shown as a function of the anisotropy in Figure 5
for spherocylinders and in Figure 6 for ellipsoids. 
They are compared with simulation data of Bolhuis and Frenkel~\cite{bolhuis}, 
Frenkel and Mulder ~\cite{frenkell}, and Camp \emph{et al.}~\cite{camp}. 

\begin{figure}
{\centering \resizebox*{0.6\textwidth}{!}{\includegraphics{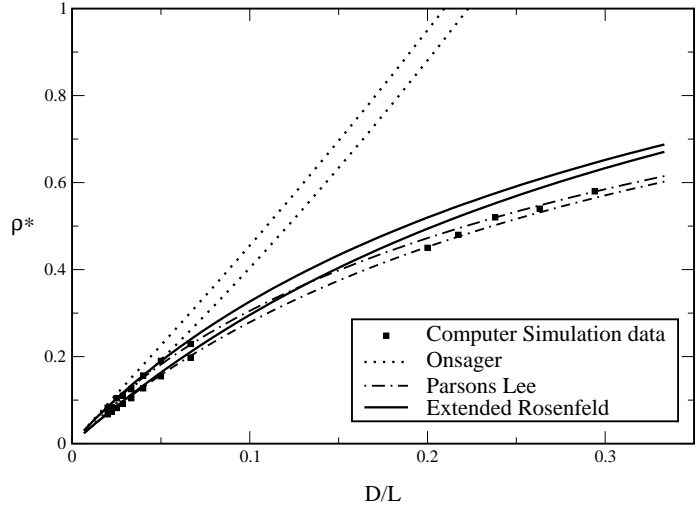}} \par}

\caption{{\small Isotropic-Nematic transition line of hard spherocylinders
as a function of elongation \protect\( D/L\protect \) and reduced
density \protect\( \rho ^{*}=\rho /\rho _{cp}\protect \) 
(\protect\( \rho _{cp}\protect \)
is the density of a close-packed lattice). Solid lines show prediction
of our approach (\ref{final}), dotted line the Onsager result (\ref{onsager}),
and dashed line the Parsons-Lee prediction (\ref{parsons}). Filled
squares are simulation results from Bolhuis and Frenkel~\cite{bolhuis}
(for larger \protect\( D/L\protect \)), only one transition density
 is given in this reference).
}}
\end{figure}
\begin{figure}
{\centering \resizebox*{0.6\textwidth}{!}{\includegraphics{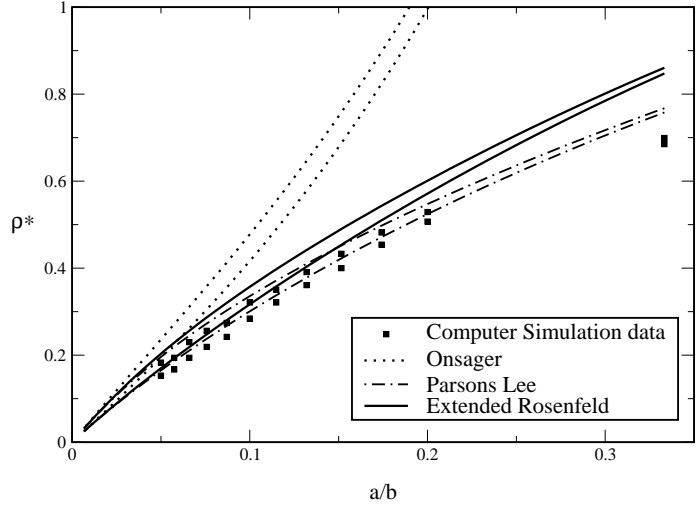}} \par}

\caption{{\small Isotropic-Nematic transition line of hard ellipsoids
as a function of elongation \protect\( a/b\protect \) and reduced
density \protect\( \rho ^{*}=\rho /\rho _{cp}\protect \) (\protect\( \rho _{cp}\protect \)
is the density of a close-packed lattice). Solid lines show prediction
of our approach (\ref{final}), dotted line the Onsager result (\ref{onsager}),
and dashed line the Parsons-Lee prediction (\ref{parsons}). Filled
squares are simulation results from Camp} \emph{\small et al.}{\small ~\cite{camp}.
The points at \protect\( k=3.0\protect \) are taken from Ref. ~\cite{frenkell}}}
\end{figure}

The figures demonstrate that our functional performs much better than
the Onsager functional, and only slightly worse than the Parsons-Lee
functional. As has already been demonstrated elsewhere for spherocylinders 
and ellipsoids~\cite{mcgrother,camp}, the predictions of the latter are 
almost exact. The slightly inferior performance of our functional is
probably related to the fact that the Parsons-Lee approach is based on 
the Carnahan-Starling equation of state, whereas the Rosenfeld
functional yields the Percus-Yevick equation of state for hard spheres, 
which is slightly inferior. On principle, one can modify the Rosenfeld
functional such that it reproduces the Carnahan-Starling equation of 
state for fluids, along the lines of an approach suggested by 
Tarazona~\cite{tarazona3}. Unfortunately, this is done at the expense 
of a less accurate description of crystals~\cite{tarazona3}. 

Our approach could possibly be improved by making the coefficient of 
$\Phi_3$ in Eqn. (\ref{8.3bis}) dependent on the orientation distribution 
$f(\varpi)$: formally, the Rosenfeld functional has the form of a third order 
$Y$-expansion~\cite{Y}. Mulder and Frenkel ~\cite {YA} and Tjipto-Margo 
and Evans~\cite{tiptomargo} have generalized the latter to convex anisotropic 
bodies and applied it to hard ellipsoids, determining the third virial 
coefficient numerically. Their nematic-isotropic coexistence densities 
were lower than those observed in the simulations. In contrast, our 
functional tends to overestimate the coexistence densities. This suggests 
that introducing orientation dependent coefficients will shift the
transition lines in the correct direction.

However, these modifications complicate the functional. We have shown
that already our simple version describes uniform fluids reasonably well.
Compared to the Parsons-Lee theory, it has the advantage of being
based on a reference density functional which describes the local 
structure of hard spheres very accurately. It interpolates on the level 
of the direct correlation function, i. e. the local bulk structure, 
between the Percus-Yevick solution for hard spheres, which is very good, 
and the Onsager solution for infinitely elongated particles, which is exact.
Therefore, we believe that it will be suited to describe inhomogeneous 
anisotropic fluids. 

\section{Summary and Outlook}

We have introduced a new density functional for liquid crystals, which
interpolates between the successful Rosenfeld functional for
hard sphere fluids~\cite{rosenfeld1} and the Onsager functional for liquid
crystals~\cite{onsager}. As a first test, we have calculated the
nematic-isotropic phase diagram for hard spherocylinders and hard
ellipsoids and shown that the new functional produces reasonable results.
In the next step, it shall be applied to calculate the local liquid
structure in homogeneous fluids. For example, direct correlation functions
in nematic fluids of ellipsoids have been determined recently from
computer simulations~\cite{phuong2,phuong3,phuong4}. They can also
be calculated from our density functional. This will be a second,
much more sensitive test of the theory. 

If the new functional is successful, it will allow to predict the
structure and the properties of nematic liquid crystals in the vicinity
of inhomogeneities. It will thus contribute to an improved microscopic
understanding of surface phenomena such as surface anchoring on rough
and structured surfaces, or of defect structures and defect interactions. 
It also provides a new approach to smectic ordering, and might lead to 
useful insight into the nature of the nematic-smectic transition~\cite{smectic}.

\section*{Acknowledgments}

We thank Guido Germano for fruitful interactions. We are grateful to
Yasha Rosenfeld for useful comments on the manuscript and for pointing 
out reference~\cite{smectic} to us. This work was funded by the 
Scuola Normale Superiore di Pisa and by the German Science 
Foundation.


\begin{thebibliography}{10}
\bibitem{note} 
The direct correlation functions in the Parsons-Lee model are directly 
proportional to the Mayer-function at all densities.
\bibitem{note2}
  Let \( \kappa _{a}({\mathbf{r}}) \) and \( \kappa _{b}({\mathbf{r}}) \)
  be the two local principal curvatures at the surface point 
  \( {\mathbf{r}} \). Then the mean curvature is 
  \( H_{i}({\mathbf{r}})=(\kappa _{a}+\kappa _{b})/2 \),
  and the Gaussian curvature, 
  \( K_{i}({\mathbf{r}})=\kappa _{a}\kappa _{b} \).
\bibitem{footnote} Our functional can be cast in the form of a 
  weighted density functional if the weighted densities are 
  allowed to depend on orientations.
\end{thebibliography}
\end{document}